\documentclass[aps,showpacs,prep,graphics,twocolumn]{revtex4}

\usepackage{amssymb}
\usepackage[dvips]{graphicx}
\usepackage{amsmath}
\usepackage{bm}
\usepackage{epsfig}
\usepackage{graphicx}
\usepackage{caption}
\usepackage{wrapfig}
\usepackage{color}

\begin{document}

\title{Extending $\Lambda(t)-$CDM to the inflationary epoch using  dynamical foliations and a pre-inflationary vacuum energy from 5D geometrical vacuum as a unifying mechanism}

\author{ $^{1}$ Jos\'e Edgar Madriz Aguilar\thanks{E-mail address: madriz@mdp.edu.ar}, $^{2}$ J. Zamarripa, $^{1}$ M. Montes, $^{1}$ J. A. Licea, $^{2}$ C. De Loza and $^{3}$ A. Peraza} 
\affiliation{$^{1}$ Departamento de Matem\'aticas, Centro Universitario de Ciencias Exactas e ingenier\'{i}as (CUCEI),
Universidad de Guadalajara (UdG), Av. Revoluci\'on 1500 S.R. 44430, Guadalajara, Jalisco, M\'exico.\\
\\
$^{2}$ Centro Universitario de los Valles, Carretera Guadalajara-Ameca Km 45.5, C. P. 46600, Ameca, Jalisco, M\'exico.\\
and\\
$^{3}$ Departamento de F\'isica,\\
Centro Universitario de Ciencias Exactas e Ingenier\'ias (CUCEI),\\
Universidad de Guadalajara.\\
Av. Revoluci\'on 1500 S. R. 44430,\\
Guadalajara, Jalisco, M\'exico.\\
E-mail: jose.madriz@academicos.udg.mx, jose.zamarripa@academicos.udg.mx, mariana.montes@academicos.udg.mx, antonio.licea@academicos.udg.mx, cynthia.deloza2536@alumnos.udg.mx,  americo.peraza@academicos.udg.mx }

\begin{abstract}
	
In this letter assuming a 5D quantum pre-inflationary vacuum energy,  we propose  a manner to extend some $\Lambda(t)$-CDM models to the inflationary period by using dynamical foliations of the five-dimensional (5D) Ricci-flat space-time manifold, regarding a non-compact extra space-like  coor\-dinate. In this formalism we achieve also a geometrical unification of inflation and the present acce\-le\-rating epoch. In this approach inflation is generated by a pre-inflationary quantum vacuum energy that maintains the 5D classical vacuum on cosmological scales. We obtain from geometrical conditions that we can model the presence of the pre-inflationary vacuum energy in 4D as a dynamical cosmological constant. The 4D inflationary period results to be governed by a power law expansion and for certain values of some parameters of the model we obtain an spectral index satisfying  $0.9607\leq n_s\leq 0.9691$ and a scalar to tensor ratio $r=0.098$, values that fit well according to Planck 2018 results. The 4D inflationary potential is induced for the 5D geometry and the 4D pre-inflationary potential is determined by the model. We show that
in this theoretical framework the present acceleration in the expansion of the universe can be explained due to a remanent of this pre-inflationary vacuum energy scaled to the present epoch and that its description can be done with the same $\Lambda(t)$. In this period we obtain a deceleration parameter in agreement with Planck 2018 data under certain restrictions of the parameters of the model. From the geometrical point of view  $\Lambda(t)$ is depending on the dynamical foliation of the 5D space-time manifold.

\end{abstract}

\maketitle

\vskip .5cm
 Induced matter theory, dynamical cosmological constant,pre-inflation, inflation, accelerated cosmic expansion.

\section{Introduction}

The acceleration in the present expansion of the universe continues being one of greatest paradigms in cosmology \cite{R1,R2}. Many different proposal have been put forward in trying to explain the origin and dynamics of that acceleration.  One of the most popular ideas is the dark energy, which is assumed to drive the expansion of the universe on late times. However until now we cannot understand accurately its  origin and physical essence. Thus, alternative models coming from theories with extra dimensions have been also considered as possible routes. The simplest dark energy model is the known as cosmic concordance, in which the dark energy is des\-cri\-bed by a cosmological constant. Unfortunately, this feature is also its main trouble due to the cosmological constant problem \cite{R3}. The $10^{121}$ orders of magnitud of difference between the observational value of the cosmological constant and the one predicted by quantum field theory, led to the emergence of dynamical vacuum decay models, also known as dynamical cosmological constant models or $\Lambda(t)-$CDM \cite{R4}. The main idea in these models is consider a vacuum energy driven by $\Lambda(t)$, whose decaying is a function of the Hubble parameter $H(t)$.  Some of them are $\Lambda(t)=mH^2$ and $\Lambda(t)=\sigma H$ \cite{R5,R6,R7,R8}. For  the first one the spectrum of gravitational waves during inflation results to be nearly scale invariant, in good agreement with CMB observations, whereas for the second one the same spectrum fails in this characteristic \cite{R9}. However, a model $\Lambda(t)=\sigma H$ seems to work in agreement with observations in the present, radiation and matter dominated epochs \cite{R7}. An interesting analysis of the evolution of the cosmological scale factor in presence of a dynamical cosmological constant has been studied in \cite{R9A}.\\

The other period of accelerated expansion of the universe is the early inflation. It is normally believed that the expansion is governed in this period for a vacuum energy modeled by the inflaton scalar field with a dominant potential over its kinetic part (the slow-roll regimen). However, depending of the model, there is an energy gap between the energy scale when inflation starts and the Planck scale of about 3 orders of magnitude. Thus the universe could have passed for a pre-inflationary period. Pre-inflarionary scenarios have been proposed in different frameworks, for example, in big bounce models in loop quantum cosmology (LQC) \cite{R10a1,R10a2,R10a3,R10a4}, in relativistic quantum geometry approaches \cite{R10a5} and in holographic cosmology \cite{R10a6}. Observational signatures of that period are being investigated \cite{R10a7}. \\

On the other hand, physical models with extra dimensions have been a recourse to explain the  acceleration of the cosmic expansion. Some of the most popular theories with extra dimensions are string theory, brane world theories and the induced matter theory. However, mo\-dels in the framework of extra dimensions must comply some physical constraints, like for example the variation of the four-dimensional Newton constant \cite{R10}. In the induced matter theory, first proposed by Paul Wesson and collaborators, the Einstein field equations of general relativity are locally and isometrically embedded into a 5D Ricci flat space-time manifold. As a difference with respect other theories with extra dimensions, in the induced matter theory the four-dimensional Newton law of gravity is not modified \cite{R11,R12}. Besides matter appears in 4D induced by the 5D geometry. In particular, the idea of inducing a cosmological constant in the framework of the induced matter theory was first done in \cite{R13}. The idea of decaying dark energy was also explored in the context of the induced matter theory in \cite{R13,R13A}. Some other approaches like $\Lambda(t)$CDM scenarios in the light of 5D Brans-Dicke theory can be found for example in \cite{R14}.\\

The plan of this letter is as follows. Section I is devoted to an introducction. In section II we develop the general geometrical formalism for the 5D pre-inflationary vacuum energy by inducing a dynamical cosmological constant employing dynamical foliations of the 5D space-time. In section III we apply our formalism to the description of the inflationary epoch. Sectio IV is left for the analysis in the present epoch of accelerated expansion. Finally in section V we give some final comments.\\

Our conventions are latin indices run from 0 to 4, with exception of $i$ and $j$ that take values from 1 to 3. Greek indices run the range from 0 to 3. The 5D metric signature we use is $(+,-,-,-,-)$. Finally we adopt units on which the speed of light $c=1$.

\section {The geometrical dynamical cosmological constant}

 Let us start assuming that our 5D universe passed for a pre-inflationary epoch  governed by a quantum vacuum energy described by a scalar field $\zeta$ with non-canonical kinetic term, whose action reads
\begin{equation}\label{qcu1}
{\cal S} = \int d^{5}y\,\sqrt{g_5}\left[\frac{^{(5)}R}{2\kappa_5}+\frac{1}{2}\Omega(\zeta)g^{ab}\zeta_{,a}\zeta_{,b}-V_{pi}(\zeta)\right],
\end{equation}  
where $V_{pi}(\zeta)$ is the potential associated to the pre-inflationary scalar field $\zeta$, $\kappa_5$ is the 5D gravitational coupling,  $g_5$ denotes the determinant of the 5D metric 
\begin{equation}\label{qcu2}
ds_5^2=g_{ab}\,dy^{a}dy^{b},
\end{equation} 
$^{(5)}R$ is the 5D scalar curvature defined by $^{(5)}R=g^{ab}\,^{(5)}\!R_{ab}$ being $^{(5)}R_{ab}$ the 5D Ricci tensor which satisfies $<\,^{(5)}\hat{R}_{ab}>=\,^{(5)}R_{ab}$, with $<\,>$ denoting the expectation value and $\hat{R}_{ab}$ is the Ricci curvature operator related to the coordinate operator \cite{newr1}
\begin{eqnarray}
\hat{y}^{a}(y^b)&=&\frac{1}{(2\pi)^{3/2}}\int d^3k_r\,dk_l \,\hat{E}^{a}\left[b_{k_rk_l}\Theta_{k_rk_l}(y^{b})\right.\nonumber\\
&& + \left.  b_{k_rk_l}^{\dagger}\Theta_{k_rk_l}^{*}(y^{b})\right], \label{qcu3}
\end{eqnarray}
with $b_{k_rk_l}^{\dagger}$ and $b_{k_rk_l}$ being the creation and annihilation operator of the 5D space-time determined by the relation
\begin{equation}\label{qcu4}
\left<\Psi\left|\left[b_{k_rk_l},b^{\dagger}_{k_rk_l}\right]\right|\Psi\right>=\delta^{(3)}(\bar{k}_r-\bar{k}_r^{\prime})\delta(k_l-k_l^{\prime}), 
\end{equation}
where $|\Psi>$ denotes a background quantum state and $\hat{E}^{a}=\in^{a}\,_{bcdf}E^{b}E^{c}E^{d}E^{f}$ . With the help of \eqref{qcu3} it can be shown that the eigenvalues obtained from aplying the coordinate operator to a background state $\left|\Psi\right>$ satisfy
\begin{equation}\label{qcu5}
d\hat{y}^{a}\left|\Psi\right>=\,^{(5)}U^{a}ds_5\left|\Psi\right>=dy^{a}\left|\Psi\right>,
\end{equation}
being $^{(5)}U^{a}$ the 5-velocity. Thus the 5D line element satisfy
\begin{equation}\label{qcux}
\left<\Psi\left|d\hat{y}_{a}d\hat{y}^{a}\right|\Psi^{\prime}\right> =\,^{(5)}U_b\,^{(5)}U^{b}\,ds_5^2\left<\Psi|\Psi^{\prime}\right> =ds_5^2\delta_{\Psi\Psi^{\prime}}.
\end{equation}
On the other hand, according to the Wesson's induced matter theory of gravity the classical field equations are: $^{(5)}R_{ab}=0$ and therefore $\left<\hat{R}_{ab}\right>=0$. In this manner, the main idea in this paper is to consider that the quantum vacuum energy coming from a pre-inflationary stage generates the inflationary expansion maintaining at the same time the 5D vacuum on large scales. Thus we can interpret that as inflation goes on, the vacuum energy plays the role of a dynamical cosmological constant $\Lambda(t)$. As we shall show, in this formalism $\Lambda(t)$ can be expressed in terms of geometrical quantities or equivalently in terms of a physical scalar field induced on our 4D space-time by the 5D geometry. In simple terms the idea is that the pre-inflationary vacuum energy can be modeled, at the same time, by a physical scalar field and by a dynamical cosmological constant of geometrical nature.
In fact, its geometrical origin allow us to postulate that a remanent of the same vacuum energy can be used to explain the present acceleration in the expansion of the universe.\\

With the idea that we have a 5D classical vacuum on cosmological scales, the field equation for $\zeta$ derived from \eqref{qcu1} reads
\begin{equation}\label{qna1}
\Omega(\zeta)\,^{(5)}\Box \zeta+\frac{1}{2}\Omega^{\prime}g^{ab}\zeta_{,a}\zeta_{,b}+V_{pi}^{\prime}(\zeta)=0.
\end{equation}
In general the semiclassical approximation allows to write  $\zeta(y^{a})=\zeta_{c}(t)+\delta\zeta$, with $\zeta_c$ being the pre-inflationary scalar field on cosmological scales, $\delta\zeta$ denoting quantum fluctuations on small non-cosmological scales such that $<\delta\zeta>=0$. However,  remembering that the pre-inflationary scalar field $\zeta$ is in this approach considered only of quantum nature then $\zeta_c=0$. Thus we obtain the relations
\begin{eqnarray}\label{qna2}
\Omega(\zeta)\simeq \Omega(0)+\Omega^{\prime}(0)\zeta ,\\
\label{qna3}
\Omega^{\prime}(\zeta)\simeq \Omega^{\prime}(0)+\Omega^{\prime\prime}(0)\zeta,
\end{eqnarray}
where we have regarded that the quantum fluctuations are small deviations from the classical value of the field. Hence, considering these conditions it comes from \eqref{qna1} that
\begin{equation}\label{qna4} 
^{(5)}\Box\zeta+\frac{V_{pi}^{\prime}(0)}{\Omega(0)}+\frac{V_{pi}^{\prime\prime}(0)}{\Omega(0)}\zeta=0.
\end{equation}
If we restrict our model to the case where $V_{pi}^{\prime}(0)=0$ and $\Omega(0)\neq 0$ the equation \eqref{qna4} reduces to
\begin{equation}\label{qna5}
^{(5)}\Box\zeta+\frac{V_{pi}^{\prime\prime}(0)}{\Omega(0)}\zeta=0
\end{equation}
If we consider a 5D warped product manifold space-time with line element given by
\begin{equation}\label{qna6}
ds_5^2=e^{2A(l)}g_{\alpha\beta}(x)dx^{\alpha}dx^{\beta}-dl^2,
\end{equation}
with $A(l)$ being the warping factor, we obtain from \eqref{qna5} that the variation of the pre-inflationary field $\zeta$ along the fifth non-compact extra coordinate $l$, for a separable field $\zeta(x^{\alpha},l)=Q(x^{\alpha})\Delta(l)$, is governed by the equation
\begin{equation}\label{qna7}
e^{-2A}\frac{d}{dl}\left(e^{4A}\,\frac{d\Delta}{dl}\right)+\left(\frac{V_{pi}^{\prime\prime}(0)}{\Omega(0)}e^{2A}+\alpha_{l}\right)\Delta=0,
\end{equation}
where $\alpha_l$ is a separation constant.\\

 Now, to relate the pre-inflationary vacuum energy with a dynamical cosmological constant of geometrical origin, we proceed as follows. A well-known Ricci-flat solution of the field equations: $^{(5)}R_{ab}=0$, is given by the 5D line element \cite{R14,Bel1}
\begin{equation}\label{eq1}
dS_{5}^2=l^2\frac{\Lambda(t)}{3}dt^2-l^2e^{2\int \sqrt{\frac{\Lambda(t)}{3}}dt}\delta_{ij}dx^{i}dx^{j}-dl^2,
\end{equation}
where  $t$ is the cosmic time and $\Lambda(t)$ is a metric function. \\

Now, to pass from 5D to 4D we assume that the 5D manifold can be dynamically foliated by a family of generic hypersurfaces given by: $\Sigma:l=f(t)$. Dynamical foliations have been first introduced by J. Ponce de Leon in \cite{JPonce,JPonce1}.
The differential line element induced on every leaf $\Sigma$ reads
\begin{equation}\label{eq2}
dS_{\Sigma}^2=\left[\frac{1}{3}f^{2}\Lambda(t)-\dot{f}^2\right]dt^2-f^2e^{2\int \sqrt{\frac{\Lambda(t)}{3}}dt}\delta_{ij}dx^{i}dx^{j},
\end{equation}
where the dot is denoting time derivative. In order to every leaf to describe a FRW universe, the conditions
\begin{eqnarray}\label{eq3}
&& \frac{1}{3}f^2\Lambda(t)-\dot{f}^2=1,\\
\label{eq4}
&& f^2e^{2\int \sqrt{\frac{\Lambda(t)}{3}}dt}=a^2(t),
\end{eqnarray}
must be valid, where $a(t)$ is the cosmic scale factor. The 4D Einstein equations induced on every $\Sigma$ considering \eqref{eq2} and \eqref{eq3} can be written as \cite{PCOS}
\begin{eqnarray}\label{eq5}
3H^2 &=& 8\pi G \rho_{(IM)}+\Lambda(t),\\
\label{eq6}
 2\frac{\ddot{a}}{a}+H^2 &=& -8\pi G p_{(IM)}+\Lambda(t).
\end{eqnarray}
where $H=\dot{a}/a$ denotes the Hubble parameter, and the energy denstity and pressure for induced matter are given respectively by
\begin{eqnarray}\label{eq7}
\rho_{(IM)} &=& 3\left(\frac{\dot{f}}{f}\right)^2+2\frac{\dot{f}}{f}\sqrt{3\Lambda},\\
\label{eq8}
p_{(IM)} &=& -2\frac{\ddot{f}}{f}-2\sqrt{3\Lambda}\frac{\dot{f}}{f}-\frac{\dot{\Lambda}}{\sqrt{3\Lambda}}-\left(\frac{\dot{f}}{f}\right)^2.
\end{eqnarray}
The equations \eqref{eq5} and \eqref{eq6} indicate that the metric function $\Lambda(t)$ can be interpreted as a dynamical cosmological constant describing a dynamical vacuum energy. It follows from \eqref{eq3} that \cite{PCOS}
\begin{equation}\label{eqq8}
\Lambda(t)=\frac{3}{f^2}\left(1+\dot{f}^2\right),
\end{equation}
indicating that $\Lambda(t)$ is, in the geometrical sense, depending of the dynamical foliation $l=f(t)$.\\

On the other hand, as it was shown in \cite{MP} the $\Lambda(t)$-CDM model introduced in \cite{Jairso} in which $\Lambda(t)=\sigma H(t)$, with $\sigma^{1/3}\simeq 150 \,MeV$ being the energy scale of the chiral phase transitions of QCD, the spectral index for gravitational waves during inflation results to be $n_{gw}\simeq -0.5826$ which enters in contradiction with observations that indicate $n=0.9649\pm 0.0042$ \cite{OBS1}. However, as suggested in \cite{R7,HH2}, this  $\Lambda(t)$-CDM model exhibits observational viability during the radiation and matter dominated epochs as well as the present epoch. Thus, to include the inflationary epoch in a $\Lambda(t)-$CDM scenario  we propose the more general form  
\begin{equation}\label{eq9}
\Lambda(t)=(1-\eta_{sr})mH^2+\eta_{sr}\sigma H,
\end{equation}  
with $m$ being a parameter and $\eta_{sr}=-\dot{H}/H^2$ is one of the slow-roll parameters employed in usual inflationary scenarios. As it is well-known $\eta_{sr}$ is an increasing pa\-ra\-me\-ter varying from $0$ to $1$ during inflation, reaching $1$ at the end of that period.Therefore, during inflation the $H^2$ term dominates and by the end of inflation and further the $H$ term is dominant.  By equating \eqref{eqq8} with \eqref{eq9} we obtain
\begin{equation}\label{eqq9}
(1-\eta_{sr})mH^2+\eta_{sr}\sigma H=\frac{3}{f^2}\left(1+\dot{f}^2\right).
\end{equation}
Thus, it is not difficult to see that during inflation the equation \eqref{eqq9} can be approximated by
\begin{equation}\label{eq10}
(1-\eta_{sr})mH^2=\frac{3}{f^2}\left(1+\dot{f}^2\right),
\end{equation} 
while in the rest of the epochs in the evolution of the universe we have
\begin{equation}\label{eq11}
\sigma H=\frac{3}{f^2}\left(1+\dot{f}^2\right).
\end{equation}
According to the definition of $\eta_{sr}$ the equation \eqref{eq10} can be explicitly written as
\begin{equation}\label{eq12}
\left(1+\frac{\dot{H}}{H^2}\right)mH^2=\frac{3}{f^2}\left(1+\dot{f}^2\right).
\end{equation}
On the other hand, as $l$ is a noncompact spacelike coordinate, we will associate $l$ with the size of the observable universe and thus we consider $l(t)=\lambda H^{-1}$ where $\lambda$ is a parameter taking values depending on the epoch of the universe. 

\section{dynamical vacuum during inflation}

Once we have introduced the general elements of the model we need, we are now in position to obtain solutions for the inflationary epoch.\\

It follows from \eqref{eq9} that during inflation the dynami\-cal vacuum represented by $\Lambda(t)$ can be approximated by $\Lambda(t)=(1-\eta_{sr})mH^2$. The foliation function $f$ that induces such $\Lambda(t)$ is determined by \eqref{eq12}, which given that $f(t)=\lambda H^{-1}$  becomes
\begin{equation}\label{eq13}
\left(1+\frac{\dot{H}}{H^2}\right)mH^2=\frac{3H^2}{\lambda^2}\left[1+\frac{\lambda^2 \dot{H}^2}{H^4}\right].
\end{equation}
Solving \eqref{eq13}, for $\lambda^2>(3/m)$ the physical solution reads
\begin{equation}\label{eq14}
H(t)=\frac{6\lambda}{\gamma\left(t-t_i+\frac{6\lambda}{\gamma H_i}\right)},
\end{equation}
with $\gamma=\sqrt{m^2\lambda+12m\lambda-36}-m\lambda$ and where we have imposed the initial condition $H(t_i)=H_i$, being $t_{i}$ the time when inflation begins and $H_i$ the value of the Hubble parameter at that time.\\

The scale factor corresponding to the Hubble parameter \eqref{eq14} is then
\begin{equation}\label{scf1}
a(t)=a_{i}\left[\frac{\gamma H_i}{6\lambda}\left(t-t_i\right)+1\right]^{\frac{6\lambda}{\gamma}},
\end{equation}
where we have used the initial condition $a(t_i)=a_i$.\\

It is not difficult to see that the Hubble parameter \eqref{eq14}, which corresponds to a power law expansion, is obtained on the generic hypersurface determined by the dynamical foliation
\begin{equation}\label{eq15}
f(t)=\frac{\gamma}{6}\left(t-t_i+\frac{6\lambda}{\gamma H_i}\right).
\end{equation}
On the other hand, it follows from the action \eqref{qcu1} that the 4D action induced on $\Sigma: l=f(t)$ can be written as
\begin{equation}\label{corr1}
{\cal S}=\int d^{4}x\sqrt{-h}\left[\frac{^{(4)}R}{16\pi G}+\frac{1}{2}h^{\mu\nu}\Omega(\Phi)\Phi_{,\mu}\Phi_{,\nu}-U_{eff}(\Phi)\right],
\end{equation}
where $U_{eff}(\Phi)=U(\Phi)+U_{pi}(\Phi)$ with $U_{pi}(\Phi)$ denoting the induced 4D pre-inflationary potential, $h$ is the determinant of the 4D effective metric derived from \eqref{eq2},\eqref{eq3} and \eqref{eq4}, $\Omega(\Phi)$ is a well-behaved function of the inflaton field $\Phi$ and $U(\Phi)$ is the 4D potential induced from the 5D part of the kinetic term of $\zeta$, which is given by 
\begin{equation}\label{qcut1}
U(\Phi)=\left.\frac{1}{2}\Omega(\zeta)g^{ll}\zeta_{,l}^2\right|_{l=f(t)}=\frac{1}{2}\Omega(\Phi)\left(\frac{\Delta_{,l}}{\Delta}\right)^2_{l=f(t)}\Phi^2,
\end{equation}
where we have considered   that $\Phi(x^{\alpha})=\zeta(x^{\alpha},l)|_{l=f(t)}$ is the 4D induced scalar field.\\

The dynamical equation for the scalar field $\Phi$ derived from \eqref{corr1} is
\begin{equation}\label{corr2}
\Omega(\Phi)\Box\Phi+\frac{1}{2}\Omega^{\prime}(\Phi)\,h^{\alpha\beta}\Phi_{,\alpha}\Phi_{,\beta}+U_{eff}^{\prime}(\Phi)=0,
\end{equation}
where the prime is denoting derivative with respect to $\Phi$.
The corresponding Friedmann equation reads
\begin{equation}\label{corr3}
3H^{2}=8\pi G\left(\frac{1}{2}\Omega(\Phi)\dot{\Phi}^2+U_{eff}(\Phi)\right).
\end{equation}
The slow-roll parameters for \eqref{corr1} have the form
\begin{equation}\label{src1}
\epsilon=\frac{M_p^2}{2\Omega(\Phi)}\left(\frac{U_{eff}^{\prime}(\Phi)}{U_{eff}(\Phi)}\right),
\end{equation}
\begin{equation}\label{src2}
\eta=M_{p}^2\left(\frac{1}{\Omega(\Phi)}\frac{U_{eff}^{\prime\prime}(\Phi)}{U_{eff}(\Phi)}-\frac{\Omega^{\prime}(\Phi)}{\Omega(\Phi)}\frac{U_{eff}^{\prime}(\Phi)}{U_{eff}(\Phi)}\right),
\end{equation}
where $M_p^2=(16\pi G)^{-1}$. Now, implementing the field transformation
\begin{equation}\label{corr4}
\phi=\int \sqrt{\Omega(\Phi)} d\Phi,
\end{equation}
 and using \eqref{eq5} and \eqref{corr3} we arrive to
\begin{eqnarray}
\frac{3H^2}{8\pi G} = \rho_{(IM)}+\frac{\Lambda(t)}{8\pi G} = \frac{1}{2}\dot{\phi}^2-\frac{1}{2a^2}(\nabla\phi)^2 + V(\phi),\nonumber\\
\label{eq16}
\end{eqnarray}
where $V(\phi)=U_{eff}(\phi(\Phi))$ and in order to be in agreement with inflationary scenarios, the both $\rho_{(IM)}$ and $\Lambda(t)$ describe in this period va\-cuum energy. This equation establishes the equivalence between the geometrical and physical vacuum energy. From one side we have a physical vacuum energy modeled by the inflaton field and in the other the same vacuum energy can be described geometrically by a dynamical cosmological constant. \\

Now, we employ the semiclassical approximation for the inflaton field
\begin{equation}\label{eq17}
\phi(x^{\nu})=\varphi(t)+\delta\phi(x^{\nu}),
\end{equation}
where $\varphi(t)=<\phi(x^{\nu})>$ is the classical part of $\phi$ while its quantum fluctuations are described by $\delta\phi$, with $<,>$ denoting expectation value. Thus, with the help of \eqref{eq7} the classical part of \eqref{eq16} leads to
\begin{equation}\label{eq18}
3\left(\frac{\dot{f}}{f}\right)^2+2\frac{\dot{f}}{f}\sqrt{3\Lambda_c(t)}+\frac{\Lambda_c(t)}{8\pi G}=\frac{1}{2}\dot{\varphi}^2+V(\varphi),
\end{equation}
where $\Lambda_c(t)$ is the dynamical cosmological constant defined on cosmological scales given by \eqref{eqq8}. The corresponding quantum part of \eqref{eq16} then implies
\begin{equation}\label{eq19}
\frac{\dot{f}}{f}\frac{\delta\Lambda}{\sqrt{3\Lambda_c}}+\frac{\delta\Lambda}{8\pi G}=\dot{\varphi}\delta\dot{\phi}+\frac{1}{2}\delta\dot{\phi}^2-\frac{1}{2}(\nabla\delta\phi)^2+\sum_{n=1}^{\infty}\frac{V^{(n)}(\varphi)}{n!}\delta\phi^{n}.
\end{equation}
It follows from \eqref{eq5}, \eqref{eq6}, \eqref{eq7} and \eqref{eq8} that the effective equation of state parameter is given by \cite{PCOS}
\begin{equation}\label{eq20}
\omega_{eff}=\frac{p_{eff}}{\rho_{eff}}=-\left[1+\frac{2\frac{\ddot{f}}{f}-2\left(\frac{\dot{f}}{f}\right)^2+\frac{\dot{\Lambda}_c}{\sqrt{3\Lambda_c}}}{3\left(\frac{\dot{f}}{f}\right)^2+2\sqrt{3\Lambda_c}\left(\frac{\dot{f}}{f}\right)+\Lambda_c}\right],
\end{equation}
where 
\begin{eqnarray}\label{eqq20}
\rho_{eff}=\rho_{(IM)}+\frac{\Lambda_c}{8\pi G},\\
\label{eqq20a}
p_{eff}=p_{(IM)}-\frac{\Lambda_c}{8\pi G},
\end{eqnarray}
are the effective energy density and pressure, respectively.
It is not difficult to see that  $\omega_{eff}\simeq -1$ once the next condition is valid
\begin{equation}\label{eq21}
\frac{\dot{\Lambda}_c}{\sqrt{3\Lambda_c}}\ll 5\left(\frac{\dot{f}}{f}\right)^2+2\sqrt{3\Lambda_c}\,\frac{\dot{f}}{f}+\Lambda_c.
\end{equation}
With the help of \eqref{eqq8} and \eqref{eq15} the condition \eqref{eq21} becomes
\begin{equation}\label{eq22}
8A^2-6A\sqrt{1+A^2}+3\gg 2A\sqrt{1+A^2}, 
\end{equation}
where $A=\gamma/6$. The inequality \eqref{eq22} is satisfied for the interval: $0<\gamma\ll 9/2$. Using the definition of $\gamma$ it can be shown that $m\gg (1/243)(36\epsilon -63/4)^2$, where in view that during inflation $\lambda^2 > 3/m$, we have expressed this same condition as $\lambda^2=3\epsilon/m$ with $\epsilon >1$. These conditions are compatible with $\lambda>1$. However, the former is in fact required in order to the fifth extra dimension to be longer than the Hubble radius, explaining in this manner why it remains directly unobservable in our 4D universe. Therefore, when these last conditions are valid the equation of state results in $\omega_{eff}\simeq -1$, which is suitable to describe a slow roll inflationary period.\\

Employing \eqref{eq18}, \eqref{eqq20} and \eqref{eqq20a} we obtain
\begin{equation}\label{eq23}
\dot{\varphi}=\sqrt{(1+\omega_{i})\rho_{eff}}=\frac{\alpha_{eff}}{t-B},
\end{equation}
where $B=t_i-6\lambda/(\gamma H_i)$, $\omega_i$ is the value of $\omega_{eff}$ at the beginning of inflation,   and
\begin{equation}\label{eq24}
\alpha_{eff}=M_p\tilde{\alpha}=M_P\sqrt{6(1+\omega_{i})(2A^2-2A\sqrt{1+A^2}+1)}\,.
\end{equation}
Solving \eqref{eq23} we arrive to
\begin{equation}\label{eq25}
\varphi(t)=\varphi_{i}+\alpha_{eff}\ln\left(\frac{t-B}{t_i-B}\right),
\end{equation}
where $\varphi_{i}=\varphi(t_i)$, being $t_i$ the time when inflation begins. Now, with the help of \eqref{eq16}, \eqref{eqq20} and \eqref{eqq20a} it is not difficult to show that
\begin{equation}\label{eq26}
V(\varphi)=\frac{1}{2}(1-\omega_i)\rho_{eff}.
\end{equation}
By means of \eqref{eqq20} the former expression finally becomes
\begin{equation}\label{eq27}
V(\varphi)=\left(\frac{1-\omega_i}{1+\omega_i}\right)\frac{2\alpha_{eff}^2}{(t_f-B)^2}\,\exp\left(\frac{1}{\alpha_{eff}}(\varphi-\varphi_f)\right),
\end{equation}
where $\varphi_f=\varphi(t_f)$ with $t_f$ being the time when inflation ends. The expression \eqref{eq27} is the potential for inflation  induced from the 5D geometry in terms of the field $\varphi$. \\

However, by using \eqref{corr4} the potential \eqref{eq27} becomes
\begin{equation}\label{corr5}
U_{eff}(\Phi_c)=U_0\,\exp\left(\frac{1}{\alpha_{eff}}\int\sqrt{\Omega(\Phi_c)}d\Phi_c\right),
\end{equation}
where we have considered that $\Phi(x^{\mu})=\Phi_c(t)+\delta\Phi(x^{\mu})$ and 
\begin{equation}\label{corr6}
U_0=\left(\frac{1-\omega_i}{1+\omega_i}\right)\frac{2\alpha_{eff}^2}{(t_f-B)^2}\,\exp\left(-\frac{1}{\alpha_{eff}}\Phi_f\right),
\end{equation}
with $\Phi_f$ being the value of $\Phi$ at the end of inflation.\\

Now, using the slow-roll parameters \eqref{src1} and \eqref{src2} the spectral index $n_s=1-6\epsilon+2\eta$ and the scalar to tensor ratio $r=16\epsilon$ read respectively
\begin{eqnarray}
n_s &=& 1-\frac{3M_p^2}{\Omega}\left(\frac{U_{eff}^{\prime}}{U_{eff}}\right)+2M_p^2\left(\frac{1}{\Omega}\frac{U_{eff}^{\prime\prime}}{U_{eff}}-\frac{\Omega^{\prime}}{\Omega^2}\frac{U_{eff}^{\prime}}{U_{eff}}\right)\nonumber,
\\
\label{et2}\\
\label{et3}
r &=& \frac{8M_p^2}{\Omega}\left(\frac{U_{eff}^{\prime}}{U_{eff}}\right)^2.
\end{eqnarray}
With the help of \eqref{corr5} the expressions \eqref{et2} y \eqref{et3} become

\begin{eqnarray}\label{qc1}
n_s &=& 1-\frac{M_p^2}{\alpha_{eff}^2}-\frac{M_p^2}{\alpha_{eff}}\frac{\Omega^{\prime}}{\Omega^{3/2}},\\
\label{qc2}
r &=& \frac{8M_p^2}{\alpha_{eff}^2}.
\end{eqnarray}
Now, considering that $n_s$ is a constant, we can restrict ourselves to the case in which $\Omega^{\prime}/\Omega^{3/2}$ is also a constant. Thus we obtain for $\Omega(\Phi_c)$ the expression
\begin{equation}\label{qc3}
\Omega(\Phi_c)=\left(\frac{2}{2C-\theta_0\Phi_c}\right)^2,
\end{equation}
where $C$ is an integration constant and $\theta_0=\Omega^{\prime}/\Omega^{3/2}$. Employing \eqref{eq14} and \eqref{qc3}, the expressions \eqref{qc1} and \eqref{qc2} read 
\begin{eqnarray}\label{qc4}
n_s &=& 1-\frac{1}{\tilde{\alpha}^2}-\frac{\theta_0}{\tilde{\alpha}}M_p,\\
\label{qc5}
r &=& \frac{8}{\tilde{\alpha}^2}.
\end{eqnarray}
According to the PLANCK 2018 results (Planck TT, TE, EE, lowE+lensing+BKIS+BAO): $n_s=0.9649\pm0.0042$ and $r<0.10$ \cite{PLANCK18}. Thus it follows from \eqref{qc5} that $\tilde{\alpha}>\sqrt{80}\simeq 8.9443$. Therefore for a given $\tilde{\alpha}$ we obtain  an interval of values for $\theta_0$ that allows  \eqref{qc4} to reproduce the observational values for $n_s$. For example, for $\tilde{\alpha}=9$, the interval $0.1669\,M_p^{-1}\leq \theta_0\leq 0.2426 M_{p}^{-1}$ correspond to $0.9607\leq n_s\leq 0.9691$. For this value of $\tilde{\alpha}$ the scalar to tensor ratio results $r=0.098$. \\

Finally, by means of \eqref{corr5} and \eqref{qc3} we obtain
\begin{equation}\label{eep}
U_{eff}(\Phi_c)=\frac{U_0}{\left(2C-\theta_0\Phi_c\right)^{2/\theta_0\alpha_{eff}}}.
\end{equation}
This is the effective inflationary potential induced by the 5D geo\-metry. Hence, it follows from \eqref{qcut1}, \eqref{qc3} and the definition of $U_{eff}$ that the pre-inflationary potential is given by
\begin{eqnarray}
U_{pi}(\Phi_c) &=& \frac{U_0}{\left(2C-\theta_0\Phi_c\right)^{2/\theta_0\alpha_{eff}}}-\nonumber\\
&&\frac{1}{2}\left(\frac{2\Phi_c}{2C-\theta_0\Phi_c}\right)^2\left(\frac{\Delta_{,l}}{\Delta}\right)^2_{l=f(t)}, \label{qnnag1}
\end{eqnarray}
where according to \eqref{qna7} and \eqref{eq1} the function $\Delta(l)$ obeys the equation
\begin{equation}\label{qnnag2}
\frac{d^2\Delta}{dl^2}+\frac{4}{l}\frac{d\Delta}{d l}+\left(\frac{V_{pi}^{\prime\prime}(0)}{\Omega(0)}+\frac{\alpha_l}{l^2}\right)\Delta=0.
\end{equation}
The general solution for \eqref{qnnag2} reads
\begin{equation}\label{qnnag3}
\Delta(l)=B_1 l^{-3/2} J_{\mu}\left[Z(t)\right]+B_2 l^{-3/2} Y_{\mu}\left[Z(t)\right],
\end{equation}
with $J_{\mu}[Z(t)]$ and $Y_{\mu}[Z(t)]$ denoting the first and second kind Bessel functions, $B_1$ and $B2$ being integration constants, $\mu=(1/2)\sqrt{9-4\alpha_l}$ and where 
\begin{equation}\label{qnnag4}
Z(t)=\left(\frac{V_{pi}^{\prime\prime}(0)}{\Omega(0)}\right)^{1/2}\, l \,.
\end{equation}
Thus, with the help of \eqref{qnnag3} and \eqref{qnnag4} the preinflationary potential \eqref{qnnag1} can be specified. 

\section{The present epoch}

In order to study vacuum decay on the present epoch of accelerated expansion, we will obtain solutions of \eqref{eq11}. Thus, as we used before, if we consider $f(t)=\lambda H^{-1}$ in \eqref{eq11} we obtain 
\begin{equation}\label{eq28}
\frac{\sigma\lambda^2}{3H}=1+\frac{\lambda^2\dot{H}^2}{H^4}.
\end{equation}
Using the initial condition $H(t_0)=H_0$ with $t_0$ denoting  the present time and $H_0$ the present value for the Hubble parameter, the physical solution for \eqref{eq28} has the form
\begin{equation}\label{eq29}
H(t)=\frac{12\sigma\lambda^2}{\sigma^{2}\lambda^{2}t^2-2\sigma^{2}\lambda\mu_0^2 t+\sigma\lambda\mu_0^2+36},
\end{equation}
where $\sigma\lambda^2>3H_0$ and 
\begin{equation}\label{coace1}
\mu_0=t_0\pm \frac{6}{\sigma\lambda}\sqrt{\frac{\sigma\lambda^2}{3H_0}-1}\,.
\end{equation}
It follows from \eqref{coace1} that the scale factor is given by
\begin{equation}
a(t) = a_0\,\frac{arctanh\left[\frac{\sigma\lambda}{s_0}\left(\lambda t-\mu_0^2\right)\right]}{arctanh\left[\frac{\sigma\lambda}{s_0}\left(\lambda t_0-\mu_0^2\right)\right]}
\label{scf2}
\end{equation}
The Hubble parameter \eqref{coace1} correspond to the generic dynamical hypersurface
\begin{equation}\label{eq30}
f(t)=\frac{\sigma\lambda^2t^2-2\sigma\lambda\mu_0^2t-\sigma\lambda\mu_0^2+36}{12\sigma\lambda}
\end{equation}
Now, with the help of \eqref{eq29} the deceleration parameter $q=-a\ddot{a}/\dot{a}^2$ reads
\begin{equation}\label{eq31}
q=-\left(1-\frac{2\sigma^2\lambda^2t-2\sigma^2\lambda\mu_0^2}{12\sigma\lambda^2}\right).
\end{equation}
According to the Planck 2018 results the present values for the deceleration parameter are $q_0=-0.5581^{+0.0273}_{-0.0267}$ \cite{PLANCK18}.
Evaluating \eqref{eq31} in the present time $t_0$ and using that $\lambda>1$ we obtain
\begin{equation}\label{ptc1}
\lambda=\frac{\sigma\mu_0^2}{\sigma t_0 -6(1+q_0)}>1.
\end{equation}
This condition is valid for $\mu_0^2>t_0-\frac{6}{\sigma}(1+q_0)>0$. This last condition implies $\sigma >(6/t_0)(1+q_0)$. Thus we have for $q_0$ the interval $-0\mbox{.}5848 < q_0 <-0\mbox{.}5308$ and in this manner for $q_0=-0\mbox{.}5848$ the values of $\sigma$ satisfy $\sigma > 2.4912\,t_0^{-1}$ while for $q_0=-0\mbox{.}5308$ we obtain $\sigma > 2\mbox{.}8152\,t_0^{-1}$. Therefore \eqref{eq29} describes an scenario with an accelerating expansion determined by a deceleration parameter $q_0$ compatible with present observational values.\\

\section{Final remarks}

In this letter assuming a 5D quantum pre-inflationary vacuum energy, we have derived an inflationary and present accelerating expansion scenarios from a 5D va\-cuum employing dynamical foliations of the 5D metric. The 5D quantum pre-inflationary vacuum energy is modeled by a scalar field with non-canonical kinetic energy and a pre-inflationary potential. We show that the vacuum energy density coming from the 5D pre-inflationary period, induced on our 4D universe, can be geometrically modeled by a dyna\-mical cosmological constant determined by a family of dynamical hypersurfaces. This is due to the condition that the pre-inflationary vacuum energy is characterized by maintaining the 5D classical vacuum on cosmological scales i.e. such that $^{(5)}R_{ab}=0$. Hence, we use the Wesson's induced matter aproach to obtain a geometrical equation of state on every leaf member of the foliation to the 5D space-time. \\

On the other hand, with the indea to incorporate an inflationary scenario in a $\Lambda(t)$-CDM scenario and inspired in the fact that the model $\Lambda(t)=\sigma H(t)$ enters in contradiccion  with the value of the spectral index for gravitational waves, inferred by observations,  during inflation  \cite{MP}, we have proposed the combined dynamical cosmological constant:  $\Lambda(t)=(1-\eta_{sr})mH^2+\eta_{sr}\sigma H$ given in  \eqref{eq9}. As a result we obtain that the term $H^2$ is dominant during inflation and for the remaining epochs the $H$ term becomes relevant.\\

The 4D slow-roll inflationary  scenario is driven by a scalar field with non-canonical kinetic term whereas the inflationary potential is induced from the 5D geometry and results to be of the form shown in \eqref{eep}. We obtain that the dynamical foliation that generates the 4D inflationary model here developd \eqref{eq15} is linear with the cosmic time. The expansion of the universe during inflation is given by a power law.  We obtain that for $\tilde{\alpha}=9$, the scalar to tensor ratio results $r=0.098$. Besides, the interval $0.1669\,M_p^{-1}\leq \theta_0\leq 0.2426 M_{p}^{-1}$ correspond to $0.9607\leq n_s\leq 0.9691$. Something  interesting that we would like to stress is that not only the inflationary potential was obtained, but also the model determines the pre-inflationary potential.
 \\

On the other hand, during the present epoch we found also an exponential law for the cosmic scale factor with a deceleration parameter compatible with Max Planck 2018 results when $\sigma >(6/t_0)(1+q_0)$. Therefore 
for $q_0=-0\mbox{.}5848$ the parameter $\sigma$ satisfy $\sigma > 2.4912\,t_0^{-1}$ while for $q_0=-0\mbox{.}5308$ we obtain $\sigma > 2\mbox{.}8152\,t_0^{-1}$. In this manner we obtain an inflationary and a present accelereting expansion scenarios induced from a 5D vacuum generated by the same geometrical mechanism and in this sense we can say that it acts as a unifying mechanism.

\section*{Acknowledgments}

\noindent  J.E.Madriz-Aguilar, M. Montes and J. A. Licea   acknowledge CONACYT
M\'exico and Departamento de Matem\'aticas of Centro Universitario de Ciencias Exactas e Ingenierias (CUCEI) of Universidad de Guadalajara for financial support. J. Zamarripa and C. De Loza acknowdlege CONACYT M\'exico  and Centro Universitario de los Valles of Universidad de Guadalajara for financial support. A. Peraza acknowledges Departamento de F\'isica of CUCEI of Universidad de Guadalajara for financial support.

\end{document}